\documentclass[amsmath,amssymb,superscriptaddress,showpacs]{revtex4-2}

\usepackage{graphicx}
\usepackage[caption=false]{subfig}
\usepackage{dcolumn}
\usepackage{bm}
\usepackage[colorlinks=true,citecolor=blue,urlcolor=blue,linkcolor=red,bookmarks=true]{hyperref}
\usepackage{placeins}
\usepackage{multirow}

\begin{document}

\title{Topological Interpretations in Anisotropic Kerr-Newman-AdS Black Hole Thermodynamics }
\author{Sreeshma A.} 
\email{rs\_sreeshmaa@kannuruniv.ac.in} 
\affiliation{Department of Physics, Kannur University, Payyanur, Kerala 670327, India}
\author{Md Sabir Ali.} 
\email{alimd.sabir3@gmail.com, sabirali@mahishadalrajcollege.ac.in} 
\affiliation{Department of Physics, Mahishadal Raj College, Purba Medinipur, West Bengal 721628, India}
\author{ C.L. Ahmed Rizwan.}
\email{ahmedrizwancl@gmail.com}
\affiliation{Department of Physics, Kannur University, Payyanur, Kerala 670327, India}
\date{\today}

\begin{abstract}
The interplay between black hole thermodynamics and topology provides a strong, coordinate-independent framework for discovering the underlying structure of black hole phase transitions as well as their universal features. We explore the thermodynamic topology of the anisotropic Kerr-Newman-AdS black hole using Duan's $\phi$-mapping theory and a complex residue approach. By creating the generalized off-shell free energy, we propose a thermodynamic vector field in the parameter space with zero points corresponding to equilibrium black hole states. The winding numbers associated with these defects are utilized to calculate the black hole's global topological number. We begin with an analysis of the entire anisotropic Kerr-Newman-AdS black hole, followed by two limiting examples. Setting the anisotropic parameter $K = 0$ produces the Kerr-Newman-AdS black hole, while eliminating the AdS contribution produces the Kerr-Newman black hole. The anisotropic Kerr-Newman-AdS and Kerr-Newman-AdS black holes have a single thermodynamic defect with winding number $+1$, which results in $W = +1$. In contrast, the Kerr-Newman black hole has two defects with winding numbers $+1$ and $-1$, which result in a vanishing global topological number, $W = 0$. These findings show that the AdS contribution alters the global topological classification, but the anisotropic matter studied here preserves the topological class of the Kerr-Newman-AdS black hole. As an analogous limiting instance, we further study a Kiselev-type black hole, analyzing the circumstances in which its metric reduces to a charged black-hole geometry and how the thermodynamic topology reflects this reduction.  Finally, we use the complex residue approach on the anisotropic Kerr-Newman-AdS black hole. The winding numbers derived from the residues are consistent with those produced by Duan's $\phi$-mapping technique, providing an independent verification of the topological classification.
\end{abstract}

\maketitle

\section{Introduction}
Black holes are among Einstein's most astonishing predictions in General Relativity, and they give a great framework for studying gravity in strong fields. The LIGO-Virgo-KAGRA teams detected gravitational waves \cite{abbott2016properties,abbott2016gw151226}, and the Event Horizon Telescope directly imaged the supermassive black holes M87* and Sagittarius A* \cite{event2022first}. These landmark observations have ushered in a new era of precision gravitational physics, allowing for rigorous tests of gravitational theories under extreme conditions. Beyond their astronomical importance, black holes are excellent natural laboratories for studying the interaction of thermodynamics, quantum physics, and gravitation. As a result, one of the main goals of modern theoretical physics is to comprehend the mathematical and physical characteristics of black holes. Black holes are an essential tool for understanding the fundamental laws of nature because of their rich theoretical structure, which has established deep connections with quantum field theory, statistical mechanics, and string theory \cite{witten1991string,dabholkar2012quantum}. As a result, studying the geometrical, thermodynamic, and topological properties \cite{cai2001topology} of black holes has become one of the main goals of modern theoretical physics, offering new perspectives on gravity and the underlying structure of the Universe.

A significant advancement in black hole physics occurred with the understanding that black holes adhere to the principles of thermodynamics. Black holes have entropy proportional to the area of the event horizon and emit thermal radiation with a distinctive Hawking temperature, as demonstrated by the seminal works of Bekenstein and Hawking \cite{bekenstein1980black,crothers2020bekenstein}. This finding reveals a strong bond between statistical physics, quantum mechanics, and gravitation. The four principles of black hole thermodynamics, which closely match the classical laws of thermodynamics and offer a coherent framework for explaining the thermal behavior of black holes, were developed as a result of this discovery. Black holes show rich thermodynamic phase patterns in anti-de Sitter (AdS) spacetime \cite{papadimitriou2005thermodynamics,bizon2011weakly}, such as the well-known Hawking–Page phase transition between heat radiation and stable massive black holes \cite{li2020thermodynamics}. Davies phase transitions, which are marked by divergences in the heat capacity that indicate shifts in local thermodynamic stability, can also occur in charged and revolving black holes \cite{bhattacharya2024topological}. Our knowledge of black hole stability, critical behavior, and phase structures has been greatly improved by these amazing thermodynamic events, which have prompted in-depth investigation into their underlying microscopic and geometrical origins.

Topological techniques have become a powerful and coordinate-independent strategy for comprehending black hole thermodynamic behavior in recent years. Wei and Liu presented a unique framework that interprets black hole solutions as topological defects in an auxiliary thermodynamic parameter space, Motivated by Duan's $\phi$-mapping topological current theory \cite{wei2022black,wei2022topology}. In this method, a two-dimensional vector field is defined from an ensemble temperature, which is used to construct a generalized off-shell free energy. Thermodynamic defects are represented by the zero points of this vector field, which match on-shell black hole topology \cite{du2023topological,chu2026imprints,fairoos2024topological,duan1984structure}. The local topological properties of these defects are characterized by their winding numbers, while the sum of all winding numbers yields the global topological charge, which serves as a topological invariant of the black hole system \cite{wei2026topology,silva2026topological}. Unlike conventional thermodynamic quantities, the global topological charge remains unchanged under smooth deformations of the thermodynamic parameters, providing a universal classification of black hole phase structures. This topological framework has been successfully applied to a wide variety of black hole spacetimes and has provided valuable insights into the relationship between thermodynamic stability, phase transitions, and the underlying topological structure of gravitational systems \cite{alipour2023topological,wei2024universal}. The topological categorization of black holes has been thoroughly studied throughout a variety of known black hole solutions since its inception. These include black holes in Lovelock gravity \cite{bai2023topology}, Born–Infeld AdS black holes \cite{ali2024topology}, Einstein–Gauss–Bonnet AdS black holes \cite{liu2023topological}, vacuum and electrovacuum AdS black holes in classical general relativity \cite{yerra2022topology}, and regular AdS black holes \cite{sadeghi2023bardeen,fang2023revisiting}. Recent extensions of this topological framework to rotating black hole spacetimes \cite{wu2023topological} have created new opportunities to investigate the topological characteristics of black hole thermodynamics when angular momentum is present.

Even though the thermodynamic topology of different black hole spacetimes has advanced significantly, nothing is known about how anisotropic matter affects the topological characteristics of rotating charged black holes. The anisotropic Kerr-Newman black hole is a physically significant expansion of the classical Kerr-Newman solution, where the equation-of-state parameter ($\omega$) and the matter intensity parameter (K) characterize the surrounding anisotropic matter \cite{ashraf2026trajectories,cho2019simple}. The nature of the matter field surrounding the black hole is determined by the parameter $\omega$, which also controls the radial pressure of the anisotropic matter distribution. Different spacetime geometries and thermodynamic behaviors result from different values of $\omega$, which correlate to diverse physical circumstances. Inspired by this, we examine the thermodynamic topology of the anisotropic Kerr-Newman AdS black hole in this work for three representative values of the equation-of-state parameter ($\omega$), which span various anisotropic matter configurations ,
$\omega$ = -0.5, $\omega$ = 0, and $\omega$ = 0.5 \cite{lee2021shadow,cho2019simple}. We study how the anisotropic matter parameter affects the thermodynamic topology and phase structure of the black hole by looking at the generalized off-shell free energy, thermodynamic vector field, defect sites, winding numbers, and global topological charge for each instance. To the best of our knowledge, there hasn't yet been a comprehensive topological study of the anisotropic Kerr-Newman AdS black hole \cite{kim2020rotating} for various values of $\omega$ in the context of Duan's $\phi$-mapping topological current theory.  In this work, we systematically study the thermodynamic topology of the anisotropic Kerr–Newman AdS black hole for three different representative values of the equation-of-state parameter ($\omega$), corresponding to different anisotropic matter configurations, in order to examine how the surrounding matter field affects the locations of thermodynamic defects, winding numbers, and the global topological charge. As a result, variations in $\omega$ can significantly influence the thermodynamic stability, phase structure, and topological characteristics of the black hole. To the best of our knowledge, there has never been a thorough topological analysis of the anisotropic Kerr-Newman AdS black hole utilizing Duan's $\phi$-mapping topological current theory published in the literature \cite{zhu2025universal,rizwan2023topological,mcinnes2021magnetic,liu2025thermodynamic}. Thus, our work extends the use of topological techniques to more general spinning black hole spacetimes by offering fresh perspectives on the interaction between anisotropic matter and black hole thermodynamic topology. Thus, the current work expands the use of topological techniques to a larger class of rotating black hole solutions and offers fresh perspectives on the impact of anisotropic matter on black hole thermodynamics\cite{ahmed2021rotating}.

The significance of topological methods for describing phase structures has been brought to light by recent advances in black hole thermodynamics. Wei, Liu, and Mann created a topological description in which equilibrium black hole configurations are represented by defects in a suitably constructed thermodynamic parameter space, inspired by Duan's $\Phi$-mapping topological current  \cite{wei2022black}. In this approach, various thermodynamic phases may be categorized and their evolution throughout phase transition regimes can be studied using the topological charge linked to each defect. In this paper, we use Duan's $\phi$-mapping topological current theory to study the thermodynamic topology of the anisotropic Kerr-Newman AdS black hole. We develop the relevant thermodynamic vector field in an auxiliary parameter space after first building the generalized off-shell free energy from the black hole's thermodynamic properties. This vector field's zero points are recognized as thermodynamic defects, whose global topological charge and winding numbers are assessed to describe the black hole's topological structure. By examining three sample values, namely $\omega$ = - 0.5, $\omega$ = 0, and $\omega$ = 0.5, special attention is paid to comprehending the impact of the anisotropic matter parameter ($\omega$) on the thermodynamic topology. The impact of anisotropic matter on the black hole's topological categorization is shown by analysis and comparison of the relevant thermodynamic behavior, defect structure, and phase transitions. Lastly, the findings are examined in relation to current topological classifications and black hole thermodynamics, emphasizing the influence of anisotropic matter on the stability and phase structure of rotating charged black holes.

Complex analysis can provide a complementary viewpoint on the thermodynamic topology of the anisotropic Kerr–Newman AdS black hole \cite{fang2023revisiting}. This approach enables the representation of the off-shell thermodynamic potential as a complex function with separated poles by analytically continuing the essential thermodynamic quantities from the real horizon radius domain to the complex plane. These singularities locations and characteristics reveal important details about the underlying thermodynamic faults and the topological features that go along with them. Specifically, the corresponding residue can be evaluated using the Laurent expansion around each pole, providing a different way to characterize the local topological structure \cite{rehan2026topological}. It is practical to examine how the anisotropy parameter affects the distribution and behavior of thermodynamic defects by using this residue based approach to the anisotropic Kerr–Newman AdS geometry. The formation, annihilation, and evolution of black-hole thermodynamic defects close to critical regions can then be studied by comparison with Duan's topological current technique. The interaction between anisotropy, phase transitions, and the global topological characteristics of the black-hole thermodynamic system can be effectively investigated using this unified paradigm.

Inspired by these advances, we study the anisotropic Kerr–Newman AdS black hole's thermodynamic topology and how the anisotropy parameter affects its phase structure. Specifically, within the same theoretical framework, we examine the relevant thermodynamic configurations for a number of sample values of the parameter $\omega$. We also investigate the limiting situation $K=0$, where the anisotropic contribution vanishes, and the black hole solution reduces to standard Kerr-Newman-AdS spacetime \cite{javed2019fermions,yang2025lyapunov}. Additionally, we recover the Kerr–Newman
black hole \cite{garcia2023regular,cembranos2014kerr,zilhao2014testing,delgado2016kerr} by setting K = 0 and eliminating the AdS contribution. Then  we find the thermodynamic defects and identify their individual winding numbers using Duan's $\phi$-mapping topological current formalism, which enables us to evaluate the variation of the global topological charge with anisotropy. By analytically continuing the pertinent thermodynamic function into the complex horizon radius plane, we further utilize the complex residue approach to establish an independent characterization. The winding numbers derived from the topological current approach are then compared with the residues connected to its isolated poles. We can determine the correlation between the two complementing approaches and investigate how the anisotropic parameter influences the locations and topological characteristics of the thermodynamic defects through this combined research. The rotating Kiselev-type black hole provides a framework for investigating how surrounding anisotropic matter affects black-hole thermodynamics. The parameter $\omega$ describes the equation-of-state behavior of the surrounding matter, while K defines its contribution to spacetime geometry. For $\omega=1/3$, the matter contribution follows the same radial scaling as the electric charge term. By setting K = 0, and adopting the asymptotically flat limit $l\to\infty$, the rotating solution simplifies to the Kerr–Newman black hole. We examine the reduction of the rotating Kiselev geometry to the Kerr–Newman black hole in the proper limiting situation as a consistency check. This creates a foundation for examining the impact of anisotropic matter on the topological characteristics of rotating charged black holes and offers a standard for verifying our thermodynamic topological research.

The organization of the paper is as follows. The anisotropic Kerr-Newman-AdS black hole is introduced in Section \textcolor{red}{II}, which also provides a summary of the relevant thermodynamic quantities utilized in the analysis. Next, using Duan's $\phi$-mapping topological current framework, we construct the off-shell free energy and examine the related thermodynamic defects, giving special attention to the dependence on the anisotropy parameter $\omega$. In this section, we also look at the thermodynamic topology of the anisotropic Kerr-Newman-AdS black hole and its limiting cases, which include the Kerr-Newman-AdS, Kerr-Newman, and rotating Kiselev geometries. Section \textcolor{red}{III} presents the complex plane formulation of the thermodynamic problem. The characteristic function obtained from the off-shell free energy is analytically extended to the complex horizon radius plane, and its residues and poles are analyzed for various values of $\omega$. A comparison is made between the winding numbers derived from the topological current approach and the residue analysis. Section \textcolor{red}{IV} highlights the key findings of our investigation and addresses the physical consequences of the findings, including the impact of anisotropy on the thermodynamic topology.  

\section{The Topology of Anisotropic Kerr-Newmann-AdS Black Hole}
\label{II}
The anisotropic Kerr–Newman-AdS black hole \cite{lee2021shadow} describes a rotating and electrically charged black hole configuration embedded in an asymptotically anti-de Sitter spacetime, with an additional contribution arising from an anisotropic surrounding matter field. By adding the effects of anisotropy through the parameters K and $\omega$, where K describes the strength of the anisotropic sector and $\omega$ establishes its equation of state behavior, it expands the traditional Kerr–Newman AdS geometry. A changed horizon structure and significantly richer thermodynamic behavior result from the coexistence of angular momentum, electric charge, negative cosmological constant, and anisotropic matter. The temperature, free energy, phase structure, and thermodynamic critical behavior are all impacted by the anisotropic contribution, which specifically changes the black hole mass.
Because of these characteristics, the anisotropic Kerr-Newman AdS black hole is an ideal framework to study how topology and black-hole thermodynamics interact. The associated metric functions in the Boyer–Lindquist type coordinate representation define the geometry, and the biggest positive root of the radial function determines the horizon
\cite{ahmed2021rotating,adamo2014kerr}.

We examine a four-dimensional spacetime with a negative cosmological constant that describes a charged black hole encircled by an effective anisotropic matter field. The anisotropic matter Lagrangian is added to the equivalent Einstein–Maxwell action, which is provided by 
\begin{equation}
I=\int_{\mathcal{M}} d^4x \sqrt{-g}
\left[
\frac{1}{16\pi}
\left(R-2\Lambda-F_{\mu\nu}F^{\mu\nu}\right)
+\mathcal{L}_m
\right]+I_b .
\label{eq:action}
\end{equation}

Where $F_{\mu\nu}$ is the electromagnetic field tensor, $R$ represents the Ricci scalar, $\mathcal{L}_m$ is the effective anisotropic matter field's Lagrangian density, and $I_b$ stands for the relevant boundary terms. In this study, we explore a negative cosmic constant and use geometrized units $G=c=1$.\\
Variating the action with regard to the metric tensor yields the Einstein field equations,
\begin{equation}
G_{\mu\nu}+\Lambda g_{\mu\nu}
=8\pi\left(T_{\mu\nu}^{(\mathrm{EM})}
+T_{\mu\nu}^{(m)}\right) .
\end{equation}

For a static, spherically symmetric spacetime, we take
\begin{equation}
ds^2=-f(r)\,dt^2+\frac{dr^2}{f(r)}+r^2d\Omega^2,
\end{equation}
with an anisotropic energy-momentum tensor,
\begin{equation}
T^\mu{}_\nu
=\mathrm{diag}(-\rho,p_r,p_t,p_t),
\end{equation}
where
\begin{equation}
p_r=-\rho,\qquad p_t=w\rho.
\end{equation}
Solving the Einstein equations using this matter prescription yields \cite{wu2023topological}
\begin{equation}
f(r)=1-\frac{2M}{r}+\frac{Q^2}{r^2}
+\frac{r^2}{l^2}-\frac{K}{r^{2w}},
\end{equation}

where $M$ and $Q$ denote the mass and electric charge of the black
hole, respectively. The parameter $K$ characterizes the strength of
the anisotropic matter distribution, while $\omega$ represents its
equation-of-state parameter.

The static line element can consequently be written as
\begin{equation}
ds^2
=
-f(r)\,dt^2
+\frac{dr^2}{f(r)}
+r^2\left(d\theta^2+\sin^2\theta\,d\phi^2\right).
\label{eq:static_metric}
\end{equation}

\subsection{Boyer--Lindquist-type coordinates}

The rotating solution is then expressed in Boyer--Lindquist type
coordinates $(t,r,\theta,\phi)$. For the asymptotically AdS extension,
we introduce
\begin{equation}
\Delta_\theta
 = 
1-\frac{a^2}{L^2}\cos^2\theta,
\qquad
\Xi
=
1-\frac{a^2}{L^2} .
\label{eq:AdS_functions}
\end{equation}
$L$ stands for the AdS curvature radius, which is associated with the cosmological constant by
\begin{equation}
\Lambda = -\frac{3}{L^2}.
\label{eq:Lambda}
\end{equation}

The Boyer--Lindquist-type line element of the anisotropic
Kerr--Newman-AdS extension is written as
\begin{equation}
\begin{aligned}
ds^2={}&
-\frac{\Delta_r}{\rho^2}
\left(
dt-\frac{a\sin^2\theta}{\Xi}\,d\phi
\right)^2
+\frac{\rho^2}{\Delta_r}\,dr^2
+\frac{\rho^2}{\Delta_\theta}\,d\theta^2
\\
&+
\frac{\Delta_\theta\sin^2\theta}{\rho^2}
\left(
a\,dt-\frac{r^2+a^2}{\Xi}\,d\phi
\right)^2,
\end{aligned}
\label{eq:KNAdS_metric}
\end{equation}
where
\begin{equation}
\rho^2  = r^2+a^2\cos^2\theta ,
\label{eq:rho_AdS}
\end{equation}

The radial function for the anisotropic Kerr–Newman–AdS black hole is taken to be
\begin{equation}
{
\Delta_r
 =
(r^2+a^2)
 \left(1+\frac{r^2}{L^2}\right)
+Q^2
-2Mr
-Kr^{2(1-\omega)}
} .
\label{eq:Delta_AdS}
\end{equation}
\subsection{Kiselev to Kerr--Newman Limit}
To study the limiting behavior of rotating Kiselev-type black holes, we explore the asymptotically flat situation with a vanishing cosmological constant ($\Lambda=0$). The rotating Kiselev-type spacetime has a radial metric function as follows
\begin{equation}
\Delta_r=r^2+a^2-2Mr+Q^2-Kr^{1-3w} .
\end{equation}
The variables $M$, $a$, and $Q$ represent mass, rotation parameter, and electric charge, respectively, whereas $K$ and $w$ describe the surrounding anisotropic matter.
With the  particular choice  $w=1/3$, the term influenced by matter becomes unaffected by the radial coordinate. As a result, the radial function simplifies to
\begin{align}
\Delta_r
&= r^2+a^2-2Mr+(Q^2-K) ,
\end{align}
Introducing an effective charge parameter,
\begin{equation}
Q_{\mathrm{eff}}^2=Q^2-K ,
\end{equation}
The radial function has the following form
\begin{equation}
\Delta_r=r^2+a^2-2Mr+Q_{\mathrm{eff}}^2 ,
\end{equation}
 It has the same form as the Kerr--Newman radial function. When $\omega=1/3$, the matter term scales similarly to the charge term, and the Kerr-Newman black hole is recovered by setting K = 0 and $\l\to\infty$. Consequently, the rotating Kiselev-type metric admits an algebraic reduction to the Kerr--Newman geometry with an effective electric charge $Q_{\mathrm{eff}}$ in the limit $w=1/3$.
$\Delta_r(r_h)=0$ determines the matching horizon radius $r_+$, resulting in \begin{equation} M=\frac{r_+^2+a^2+Q_{\mathrm{eff}}^2}{2r_+} .
\end{equation}
This limiting example offers a helpful context for examining and contrasting the topological characteristics of the Kerr-Newman black hole with the spinning Kiselev-type black hole. The physical equivalency of the matter sources necessitates independent validation of the related field equations, the reduction is a geometrical equivalency at the level of the radial metric function.
\subsection{Horizon condition}

The event horizon is determined by the largest positive root of the
radial function,
\begin{equation}
\Delta_r(r_h)=0 .
\label{eq:horizon_condition}
\end{equation}

Thus, the horizon equation of anisotropic Kerr-Newman-AdS black hole  becomes
\begin{equation}
(r_h^2+a^2)
\left(1+\frac{r_h^2}{L^2}\right)
+Q^2
-2Mr_h
-Kr_h^{2(1-\omega)}
=0 .
\label{eq:horizon_equation}
\end{equation}

The mass expressed in terms of the horizon radius provides the starting
point for constructing the thermodynamic quantities and subsequently
investigating the thermodynamic topology of the black hole \cite{gao2022thermodynamics}.
Solving the horizon condition regarding the mass parameter
gives
\begin{eqnarray}
\label{mJ}
&&
M=\frac{
(r_h^2+a^2)
\left(1+\frac{r_h^2}{L^2}\right)
+Q^2
-Kr_h^{2(1-\omega)}
}{
2r_h
} ,
\end{eqnarray}
\begin{eqnarray}
\label{mK}
&&J =
\frac{a}{2r_h\Xi^2}
\left[
(r_h^2+a^2)
\left(1+\frac{r_h^2}{L^2}\right)
+Q^2-Kr_h^{2(1-\omega)}
\right] .
\end{eqnarray}
The remaining thermodynamic quantities can be expressed as
\begin{equation}
S  =
\frac{\pi(r_h^2+a^2)}
{\Xi} ,
\end{equation}
\
\begin{equation}
\Omega=
\frac{
a\left(1+\frac{r_h^2}{L^2}\right)
}{
r_h^2+a^2
} ,
\end{equation}
\
\begin{equation}
\Phi=
\frac{
Q r_h
}{
r_h^2+a^2
} .
\end{equation}
Here, $M$, $a$, $K$, and $\omega$ stand for the mass, spin parameter, strength of anisotropic matter and equation of state parameter of anisotropic matter . 
The volume conjugate to the pressure is given by  \cite{parikh2006volume}
$$V=\left(\frac{\partial M}{\partial P}\right)_{S,J,Q}$$ With the thermodynamic pressure identified with the cosmological
constant through
\begin{equation}
P = -\frac{\Lambda}{8\pi}
 = \frac{3}{8\pi L^2} .
\end{equation}
 The first
law of black-hole thermodynamics for the anisotropic
Kerr--Newman AdS black hole can then be expressed as \cite{bekenstein1980black}
\begin{equation}
dM
=
T\,dS
+\Omega_H\,dJ
+\Phi_H\,dQ
+V\,dP
+\mathcal{K}\,dK , 
\end{equation}
where $\Omega_H$ and $\Phi_H$ are the angular velocity and electric
potential at the event horizon, respectively. The term
$\mathcal{K}\,dK$ accounts for the contribution of the anisotropic
matter sector when the parameter $K$ is treated as a thermodynamic
variable.

Using the appropriate scaling properties of the thermodynamic
quantities, the corresponding Smarr relation takes the form
\begin{equation}
M
=
2TS
+2\Omega_H J
+\Phi_H Q
-2PV
+\mathcal{C}_{K}K ,
\end{equation}
where $\mathcal{C}_{K}$ represents the conjugate quantity associated
with the anisotropic parameter $K$. The precise form of
$\mathcal{C}_{K}$ depends on the thermodynamic treatment adopted for
the anisotropic matter sector.

The combined effects of rotation, electric charge, the AdS pressure, and the surrounding anisotropic matter can result in a rich phase structure in the extended phase-space thermodynamics of the anisotropic Kerr–Newman–AdS black hole. Specifically, the thermodynamic critical behavior and the corresponding defect structure can be altered by changes in the anisotropic equation of state parameter $\omega$. These characteristics offer an ideal environment for examining the topological characteristics of the black-hole thermodynamic system using both the complex residue method and Duan's $\phi$ -mapping approach.
\newline Now that we have the thermodynamic quantity formulas, we can examine the topological number of the anisotropic Kerr-Newman AdS black hole. The generalized off-shell free energy can first be introduced as described in Ref.
\cite{wei2022topology,wei2026topology}
, one can first introduce the generalized off-shell free energy as,
\begin{eqnarray}
\mathcal{F} = M -\frac{S}{\tau}, \,
\label{thermkerr}
\end{eqnarray}
$\tau$ is an auxiliary parameter that represents the inverse temperature of the cavity around a black hole. The free energy is equivalent to its on-shell value when $\tau = 1/T$.

To analyze the black hole phase transitions in terms of topological characteristics, we use Duan's $\phi$ mapping topological current theory. Ref. \cite{du2023topological,duan1984structure}, the two-component vector $\phi=(\phi^{r_h},\phi^\Theta)$ in two dimensions is conventionally written as
\begin{eqnarray}
\phi = \left(\frac{\partial \mathcal{F}}{\partial r_{h}}\, ,  -\cot\Theta\csc\Theta\right) \, .
\label{phi}
\end{eqnarray}
Here, we have assumed that the Duan's potential is the thermodynamic system's off-shell free energy. The parameter $\Theta$ ($0 \le \Theta \le +\pi$ )is used for convenience. The component $\phi^\theta$ diverges at $\theta = 0, \pi$ , with the vector pointing outward. Furthermore, as described in Ref.\cite{wei2022topology}, the extremal points of the generalized free energy landscape match perfectly to the on-shell black holes. 
Thus, the black hole solution is exactly satisfied by the zero point of the component $\phi^{r_h}$ . Eventually, $\Theta = \pi/2$  is obtained from the component $\phi^\Theta = 0$.
According to Duan's $\phi$ -mapping topological current theory, a topological current is defined as \cite{fu2000evolution,duan1998bifurcation}
\begin{eqnarray}
\label{jmu}
j^{\mu}=\frac{1}{2\pi}\epsilon^{\mu\nu\rho}\epsilon_{ab}\partial_{\nu}n^{a}\partial_{\rho}n^{b}\, , \quad
\mu,\nu,\rho = 0 ,1 ,2 .
\end{eqnarray}
where $\partial_{\nu} = \partial /\partial x^{\nu}$ and $x^{\nu}=( \tau ,~r_h ,~\Theta )$ . The normalized vector
$n = (n^r, n^\Theta)$, where $n^r  =  \phi^{r_h}/{||\phi||}$ and $n^\Theta  =  \phi^{\Theta}/{||\phi||}$
Note that the topological current (\ref{jmu}) is conserved; hence, the continuity equation $\partial_{\mu}j^{\mu} = 0$. Using the $\phi$-mapping formalism, the current can be expressed in terms of the zeros of the vector field as \cite{fu2000evolution,wei2020topological}

\begin{equation}
j^{\mu} = \delta^{2}(\phi)J^{\mu} \left(\frac{\phi}{x}\right)\ . 
\end{equation}
The three-dimensional Jacobian is represented by the quantity $J^{\mu}\left(\phi/x\right)$, which has the following form $\epsilon^{ab} J^{\mu}\left(\phi/x\right) = \epsilon^{\mu\nu\rho}\partial_{\nu}\phi^a\partial_{\rho}\phi^b$. This relationship demonstrates that the zeros of $\phi$ are where the topological current is concentrated. In other words, the topological current $j^\mu$ identically disappears except at the sites where $\phi^a(x_i)$ = 0. We express its i-th solution as x = $z_i$. The expression for the topological current density is \[
j^0 = \sum_{i=1}^{N} \beta_i\eta_i
 \delta^2\left(\vec{x}-\vec{z}_i\right)
\]
Where $\beta_i$ is the Hopf index, which represents the number of times
the mapping from the $x$-space to the $\phi$-space covers the target
space. When a closed loop in the $x$-space encircles the zero point
$\vec{z}_i$. The Hopf index is therefore a positive integer,
$\beta_i=1,2,3,\ldots$.

 As a result, we may express the topological number W in a specific parametric region $\Sigma$ as \cite{wei2022black}:
\begin{equation}
W  =  \int_{\Sigma}j^{0} d^2x =  \sum_{i=1}^{N}\beta_{i}\eta_{i} . 
\end{equation}
The value of $w_i$, which is the winding number for the $i$-th zero
point of $\phi$ contained in $\Sigma$, is independent of the
geometry of the region. Since the different zero points of the vector
field are generally isolated, the Jacobian
$J^0\left(\frac{\vec{\phi}}{\vec{x}}\right) \neq 0$,
the defect bifurcates when
$J^0\left(\frac{\vec{\phi}}{\vec{x}}\right) = 0$.
Equation~(32) demonstrates that the global topological number in any
given region is the sum of the winding numbers of all the zero points,
which represents the local property of each topological defect.
The stability of black holes is directly correlated with the winding number. A stable state of a black hole is indicated by a positive winding number ($w = +1$) , while an unstable black hole state is indicated by a negative winding number ($w = -1$) . After determining the winding number of every thermodynamic defect, the
global topological number(W) is obtained by summing all the
individual winding numbers,
\begin{equation}
W=\sum_{i} w_i.
\end{equation}
Hence, the global topological number ( $W$ ) is a defining characteristic of every black hole system and functions as a classification criterion. Because different systems have different values of topological number, these systems can be categorized according to W. So Global topological features are highlighted by this distinction.\\
\subsection{Topological classes of an Anisotropic Kerr-Newman AdS black hole}
This section discusses the anisotropic Kerr-Newman AdS black hole topology \cite{zhu2025universal}. The generalized free energy can be easily obtained from the results already provided in Eq. (\ref{thermkerr})

\begin{equation}
\mathcal{F}(r_h,\tau)
  =  
\frac{
(r_h^2+a^2)
\left[
\tau\left(9+24\pi P r_h^2\right)
-6\pi r_h\left(3-8\pi P a^2\right)
\right]
+\tau Q^2\left(3-8\pi P a^2\right)^2
-9K\tau r_h^{2(1-\omega)}
}{
2r_h\tau\left(3-8\pi P a^2\right)^2
} .
\label{offFree_E}
\end{equation}

\begin{figure}[t]
\centering

\subfloat[]{%
\includegraphics[width=0.35\linewidth]{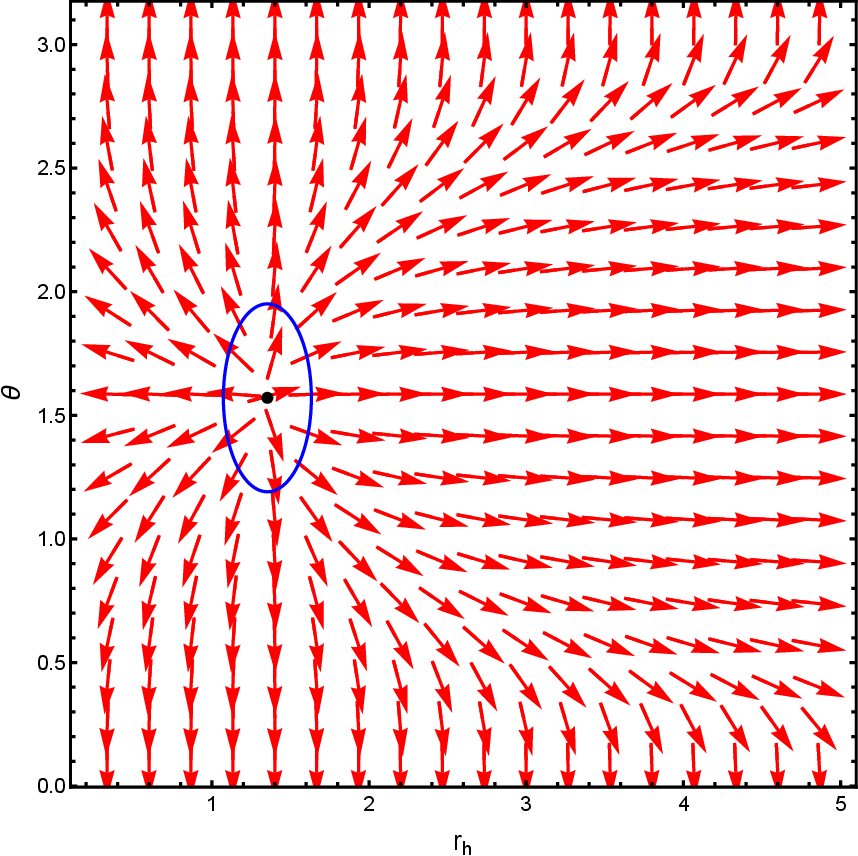}}
\hfill
\subfloat[]{%
\includegraphics[width=0.35\linewidth]{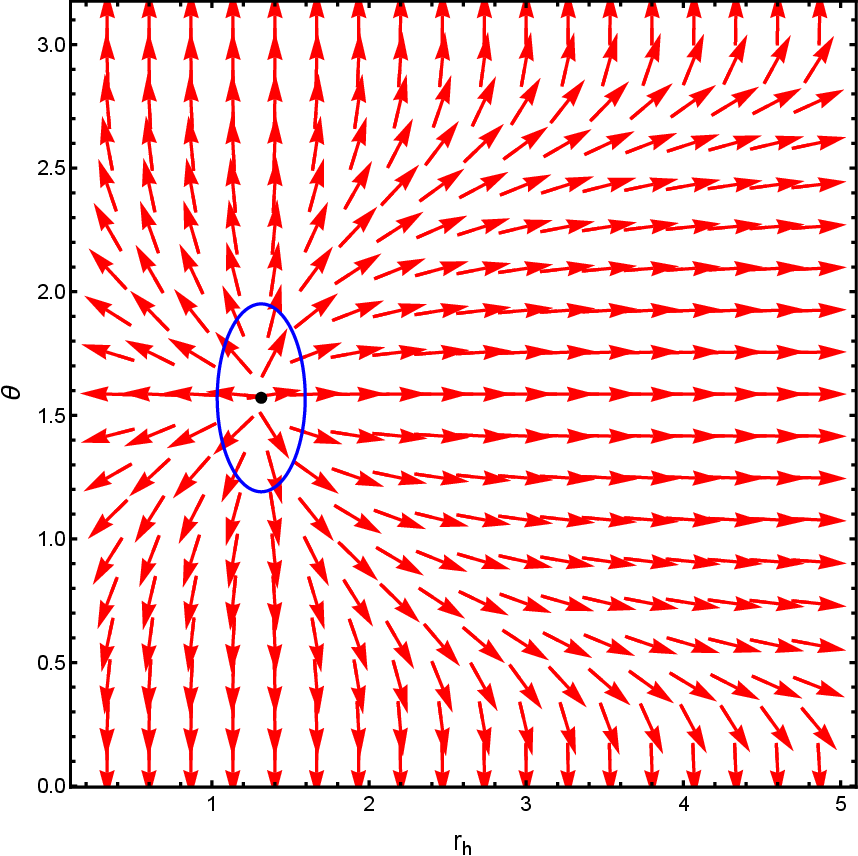}}
\hfill
\subfloat[]{%
\includegraphics[width=0.35\linewidth]{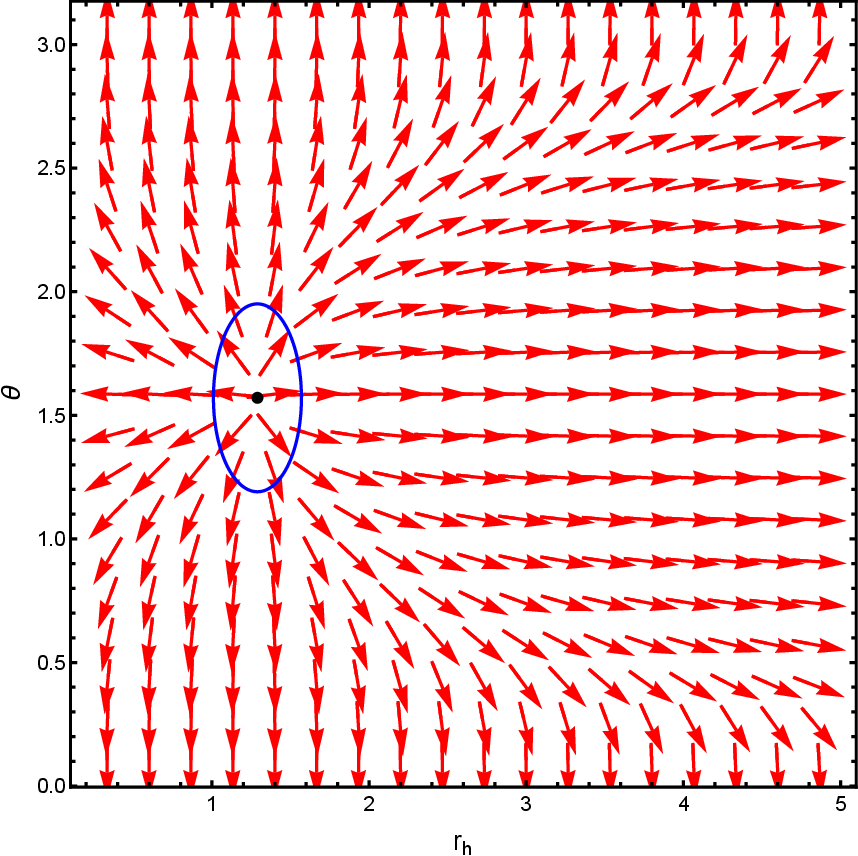}}
\caption{The way that $\phi^{r_h}$ zero points behave in the $r_h-\tau$ plane for the anisotropic Kerr-Newman AdS black hole. Panels (a), (b), and (c) are related to different values of equation of state parameter, $\omega = -0.5$, $\omega = 0$, and $\omega = 0.5$ respectively. The black dots that represent the zero points (ZPs) are situated at \((r_h,\Theta) = (1.35276,\pi/2) \) for $\omega = -0.5$, \((r_h,\Theta) = (1.3137,\pi/2)\) for $\omega = 0$ and \((r_h,\Theta) = (1.28958,\pi/2)\) for $\omega = 0.5$ . The orientation of the contour \(C\) is  anticlockwise, corresponding to the topological charge \(Q_{c} = +1\) . The total topological charge is then $Q_\text{total} = +1$ .
}
\end{figure}

\begin{figure}[t]
\centering
\includegraphics[width=0.4\textwidth]{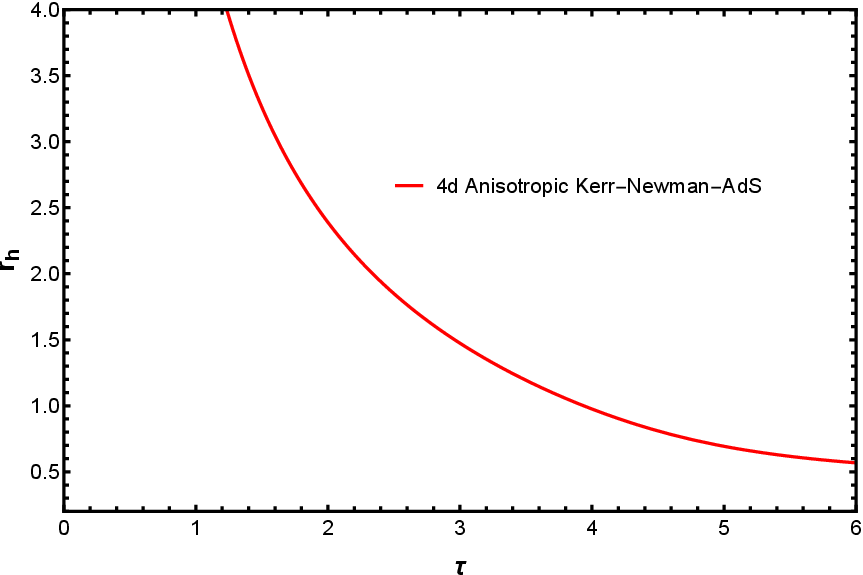}
\caption{For the anisotropic Kerr-Newman AdS black hole with $a = 0.4$, $Q = 0.2$, $K = 0.08$, and $P = 0.1193$, the plot displays the behavior of the zero points of $\phi^{r_h}$ in the $r_h-\tau$ plane. For any value of $\tau$, there is only one stable anisotropic Kerr-Newman AdS black hole.
}
\label{SchKerr}
\end{figure}

\begin{figure}[t!]
\centering
\includegraphics[width=0.4\textwidth]{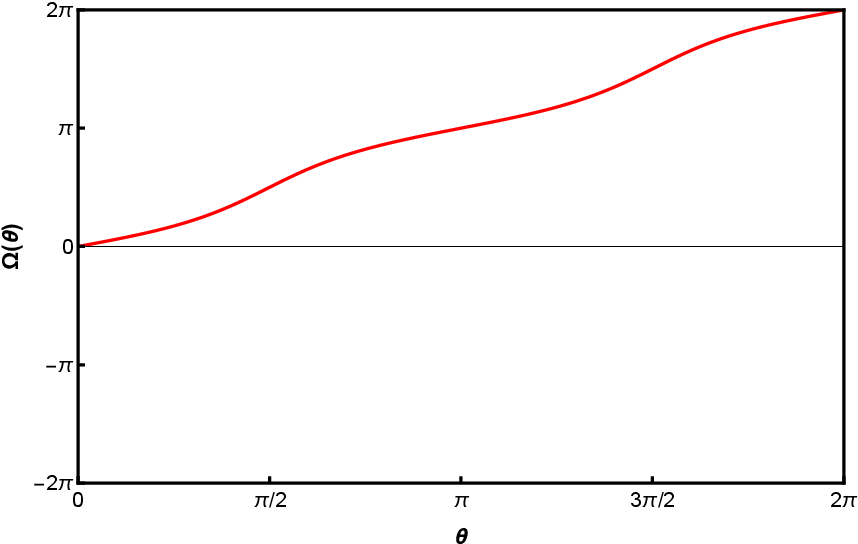}
\caption{The plot depicts the behavior of $\Omega$ versus $\vartheta$ for the contours $C$ (red solid curve) of the Anisotropic Kerr-Newman AdS black hole.
\label{Kerrsen}}
\end{figure}

\begin{figure}[t!]
\centering
\includegraphics[width=0.35\textwidth]{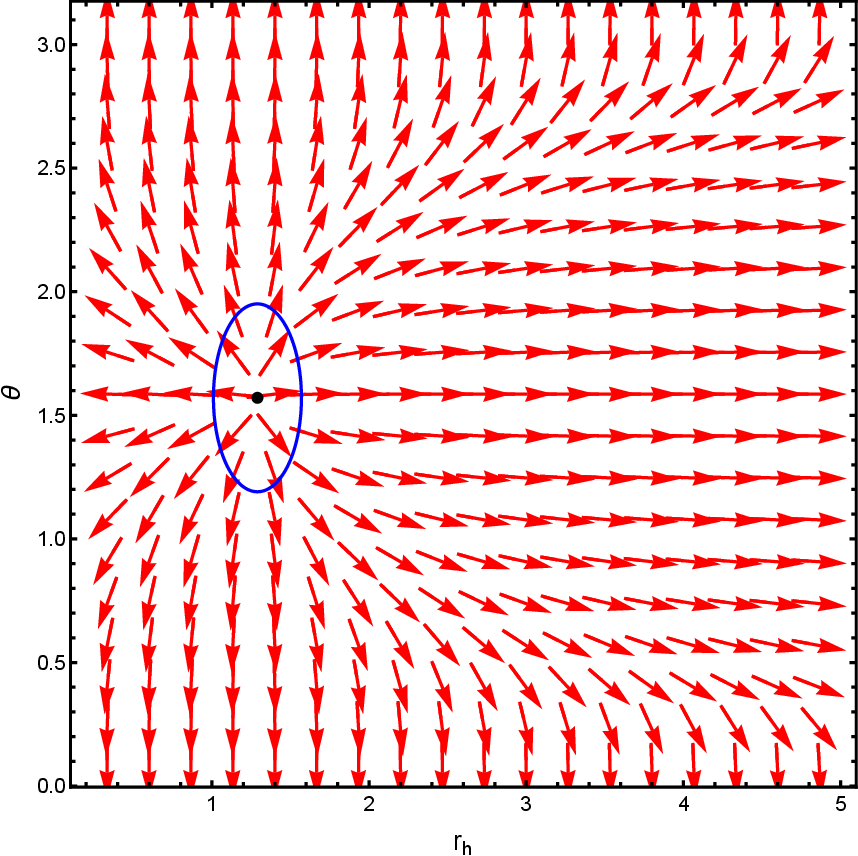}
\caption{Plotting the unit vector field $n$ on the $r_h$-$\Theta$ plane for a four-dimensional Kerr-Newman black hole in AdS spacetime with $\tau = 3.2$ when the rotation ($K = 0$) is set to zero. Black dot represent the zero point (ZP), which is situated at \((r_h,\Theta) = (1.28958,\pi/2)\) and \(C\) is counterclockwise, corresponding to the topological charges \(Q_{c} = +1\). Then the total topological charge $Q_\text{total} = +1$ .}
\label{vector0}
\end{figure}

\begin{figure}[t!]
\centering
\includegraphics[width=0.4\textwidth]{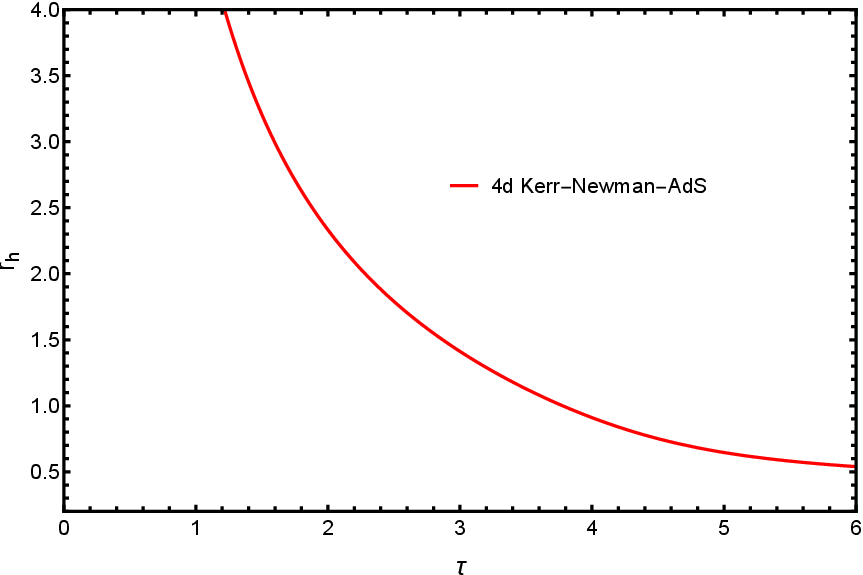}
\caption{The Plot illustrates the stable branch of the Kerr-Newman-AdS black hole phase (red curve) at a = 0.4, Q = 0.2, and P = 0.1193. \label{4dKerr_01}}
\end{figure}

\begin{figure}[t!]
\centering
\includegraphics[width=0.4\textwidth]{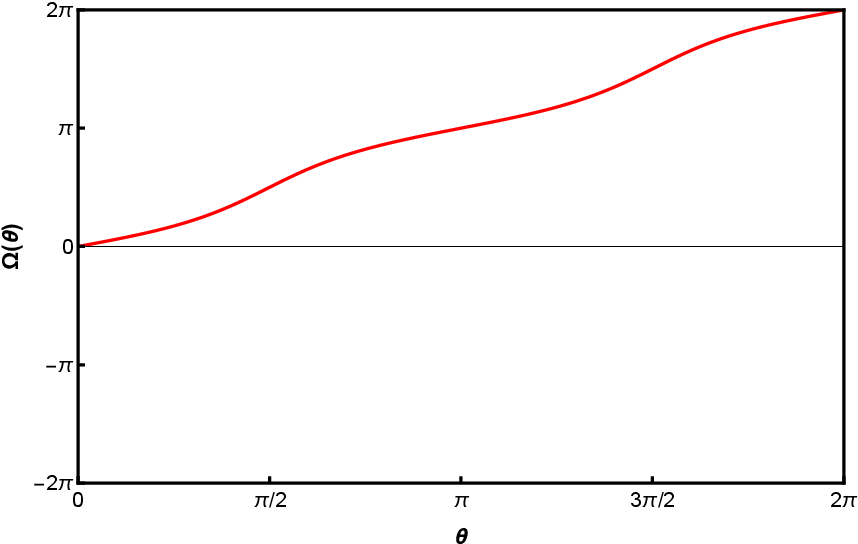}
\caption{Plot of $\Omega$ versus $\vartheta$  for contours C (red solid curve) of Kerr-Newman-AdS black hole with parameters a = 0.4, Q = 0.2, and P = 0.1193.
\label{windingspinless}}
\end{figure}
The vector $\phi$'s constituent parts can be calculated as

\begin{equation}
\phi^{r_h}
   = 
\frac{
\tau\left[
72\pi P r_h^4
+\left(9+24\pi P a^2\right)r_h^2
-9a^2
-Q^2\left(3-8\pi P a^2\right)^2
+9K(2\omega-1)r_h^{2(1-\omega)}
\right]
-12\pi\left(3-8\pi P a^2\right)r_h^3
}{
2r_h^2\tau\left(3-8\pi P a^2\right)^2
}.
\label{Phi_rh}
\end{equation}

\begin{equation}
\phi^{\Theta}  =  -\cot\Theta\,\csc\Theta .
\end{equation}
Solving the equation $\phi^{r_h}=0$ yields a relation between the horizon radius $r_h$ and the inverse temperature $\tau$, which is represented using a curve on the $r_h-\tau$ plane. Regarding the four-dimensional anisotropic Kerr-Newman-AdS black hole, this relation can be expressed as
\begin{equation}
\tau(r_h)
=
\frac{
12\pi\left(3-8\pi P a^2\right)r_h^3
}{
72\pi P r_h^4
+\left(9+24\pi P a^2\right)r_h^2
-9a^2
-Q^2\left(3-8\pi P a^2\right)^2
+9K(2\omega-1)r_h^{2(1-\omega)}
} .
\label{tau_ksads}
\end{equation}
When the anisotropic term disappears, the equations shown in Eqs.~(\ref{offFree_E}), (\ref{Phi_rh}), and (\ref{tau_ksads}) reduce to the equivalent expressions of the four-dimensional Kerr-Newman-AdS case. When a critical point is surrounded by a closed contour, the topological charge is nonzero, otherwise, it disappears. We create closed contour $C$ in order to determine the topological charge. The closed contour $C$ is defined by $\vartheta \in (0,2\pi)$ as
\begin{equation}
\begin{cases}
r  = X \cos \vartheta + r_0 \\
\theta  = Y \sin \vartheta + \dfrac{\pi}{2} .
\end{cases}
\label{contour}
\end{equation}
We select $(X ,Y ,r_0)=(0.05,0.08,1.313699)$ for contour $C$.
To better understand how the vector field behaves along each contour, We introduce a new quantity that measures the deflection of a vector field along a particular contour as
\begin{equation}
\Omega(\vartheta)=\int_{0}^{\vartheta} \epsilon_{ab}\, n^{a}\,\partial_{\vartheta} n^{b}\, d\vartheta .
\label{omega}
\end{equation}
Where $\epsilon_{ab}$ is the two-dimensional Levi-Civita symbol and $n^a$ represents the unit vector field. The contour includes the zero point for $\vartheta$ = 2$\pi$, and the topological charge is calculated by
\begin{equation}
Q=\frac{1}{2\pi}\,\Omega  (2\pi) .
\label{charge}
\end{equation}

Fig.\textcolor{red}{1} depicts the zero points associated with the vector field in the $(r_h,\Theta)$ plane for the anisotropic Kerr-Newman AdS black hole at three distinct anisotropic parameter values: $\omega$ = -0.5, $\omega$ = 0, and $\omega$ = 0.5. $a = 0.4$, $Q = 0.2$, $K = 0.08$, and $P = 0.1193$ are the remaining parameters. The vector field is shown by the red arrows. The black dot represents the zero point $\mathrm{ZP}$. The zero point is found at $(r_h,\Theta) = (1.35276,\pi/2)$ for $\omega = -0.5$, $(1.3137,\pi/2)$ for $\omega = 0$, and $(1.28958,\pi/2)$ for $\omega = 0.5$, respectively. The zero point is enclosed by the contours $C$. Consequently, $w = +1$ is the winding number for the contours $C$ (red solid). In terms of the global topological number, $W=+1$ is simply obtained. This feature is similar to the four-dimensional RN-AdS and Kerr-AdS situations, where the global topological number is $+1$ \cite{wei2022black,wu2023topological}. In this manner, we can deduce that the anisotropic Kerr-Newman AdS black hole has the same geometric topology for specific parametric values and that, in terms of thermodynamic topology, they belong to the same topological classes. The Schwarzschild-AdS black hole has a topological number of $W=0$, while the anisotropic Kerr-Newman AdS black hole has a nontrivial topological number of $W=1$. This disparity suggests that the rotation, electric charge, and anisotropic matter features have a significant impact on the topological structure of the black hole.\\
Fig.~\ref{SchKerr} shows the zero points $\phi^{r_h}$ in the $r_h-\tau$ plane for a fixed set of rotation parameter, electric charge, anisotropic parameter, equation-of-state parameter, and pressure values as $a = 0.4$, $Q = 0.2$, $K = 0.08$, $w = 0.5$, and $P = 0.1193$. As illustrated in Fig.~\ref{SchKerr}, for any value of $\tau$, there is only one thermodynamically stable anisotropic Kerr-Newman AdS black hole with these parameter values.
In Fig. \ref{Kerrsen}, we depict $\Omega(\vartheta)$ for the contour $C$. With winding number +1, the contour $C$ evolves in the positive (anticlockwise) $\Omega$ direction. The global topological number of the anisotropic Kerr-Newman-AdS black hole is $W=+1$, and it has a single zero point encircled by a closed contour. This nonzero topological charge indicates that the zero point is a topologically stable defect and demonstrates the black hole's nontrivial topological character.
Consequently, $\Omega(\vartheta)$ increases monotonically for the contour $C$ and reaches $2\pi$ at $\vartheta=2\pi$. Therefore, the critical point $\mathrm{CP}$'s topological charge is
\begin{equation}
Q_{\mathrm{CP}} =+1 .
\end{equation}
This has therefore been linked to the development of topological charges on the the contours.

The behavior of the unit vector fields $n$ in the parametric space $(r_h$, $\Theta$) for the four-dimensional Kerr-Newman black holes in AdS spacetime at inverse temperature $\tau = 3.2$ in the limit of vanishing rotation parameter ($K = 0$) is plotted in Fig.~\ref{vector0}. For $C$, we define the topological charges as \(Q_{c} = +1\), such that the sum of the total topological charge is $Q_\text{total} = +1$. The ZP contained by the contours are indicated by black dots and are located at \((r_h,\Theta) = (1.28958,\pi/2)\) for $C$. In Fig.~\ref{4dKerr_01}, we depict the zero points of $\phi^{r_h}$ in the parameter space ($r_h,\tau$) for a range of values of $a = 0.4$, $Q = 0.2$, and $P = 0.1193$ to observe how the black hole evolves. In this particular case, the Kerr-Newman-AdS black hole has a single stable black hole branch.\\

Fig.\ref{windingspinless} illustrates the variation of the winding angle
$\Omega(\vartheta)$  with respect to the parameter $\vartheta$ along the closed contour
$C$. As the contour is traversed once in the counterclockwise direction,
the winding angle increases continuously from $\Omega(0)=0$ to
$\Omega(2\pi)=2\pi$. Consequently, the total change in the winding angle
is $\Delta\Omega=2\pi$, and the corresponding winding number is +1.

\begin{figure}[t!]
\centering

\subfloat[]{%
\includegraphics[width=0.35\linewidth]{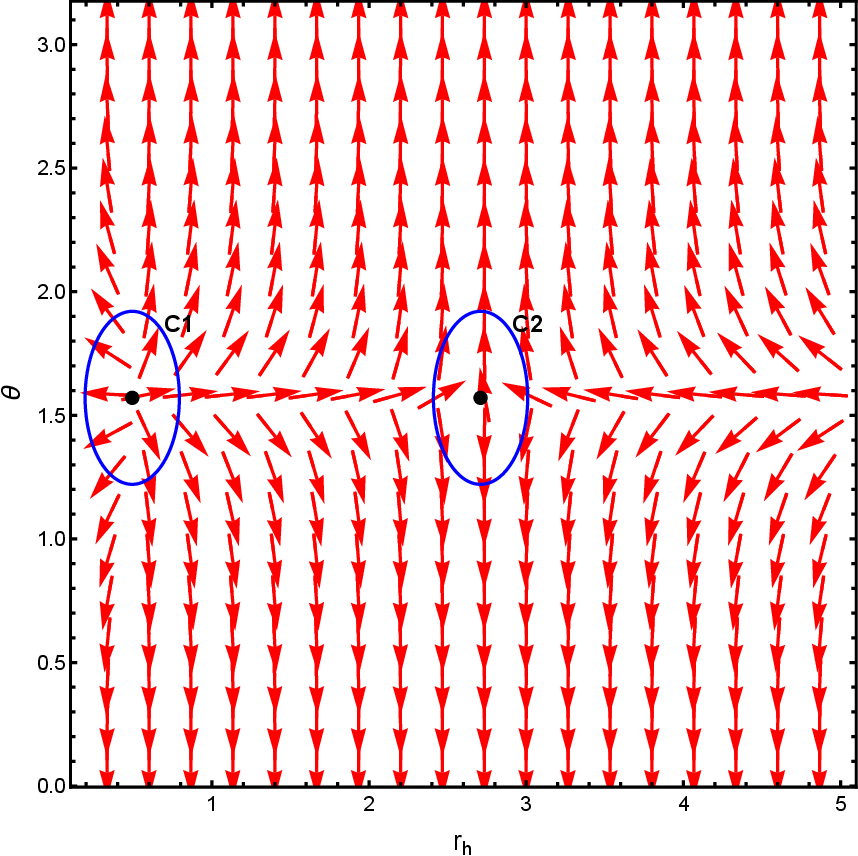}
}
\caption{A four-dimensional Kerr-Newman black hole with rotation parameter a = 0.4, electric charge Q = 0.2, and $\tau=35$ has a unit vector field $n$ on the $r_h$--$\Theta$ plane. The zero points (ZPs), indicated by black dots, are situated at
$(r_h,\Theta)=(0.492962,\pi/2)$ and $(r_h,\Theta)=(2.70932,\pi/2)$.
The contours $C_1$ and $C_2$, enclosing the zero points, carry winding number
$w_1=+1$ and $w_2=-1$, respectively.
Their sum gives the total topological winding number $W=0$.
}
\label{vectoratcritical}
\end{figure}

We then investigate the scenario in which the negative cosmological constant disappears, i.e., when $\Lambda=0$. The plot in Fig.~\ref{vectoratcritical} depicts the topological nature of the thermodynamic phase structure in the (\(r_h-\theta)\) plane of a four-dimensional Kerr-Newman black hole with the ensemble temperature $\tau=35$ in the off-shell configuration. The rotation and electric charge parameters are set at 0.4 and 0.2, respectively. The contours $C_1$ and $C_2$ are determined to be placed at $(r_h, \Theta)= (0.492962,\pi/2)$ and $(2.70932,\pi/2)$, respectively. Consider the orientations of the contours, $C_1$ and $C_2$, which are counterclockwise and clockwise respectively. The contours $C_1$ and $C_2$ correspond to the topological charges \(Q_{C_1} = +1\) and \(Q_{C_2} = -1\), respectively. This corresponds to the total topological charge, $Q_\text{total} = 0$.

\begin{table}[htbp]
\centering
\label{tab:topological_classification}

\begin{tabular}{|c|c|c|c|}
\hline
Black hole & Parameters  & Winding numbers & Total winding number \\
\hline

Kerr--Newman
& $a\neq0,\; Q\neq0,\; P=0,\; K=0$
& $+1,\;-1$
& $0$ \\
\hline

Kerr--Newman--AdS
& $a\neq0,\; Q\neq0,\; P>0,\; K=0$
& $+1$
& $+1$ \\
\hline

Anisotropic Kerr--Newman--AdS
& $a\neq0,\; Q\neq0,\; P>0,\; K\neq0$
& $+1$
& $+1$ \\
\hline

\end{tabular}
\caption{Topological classification of the Kerr--Newman,
Kerr--Newman--AdS, and anisotropic Kerr--Newman--AdS black holes.}
\end{table}

\begin{figure}[htbp]
\centering

\subfloat[]{%
\includegraphics[width=0.35\linewidth]{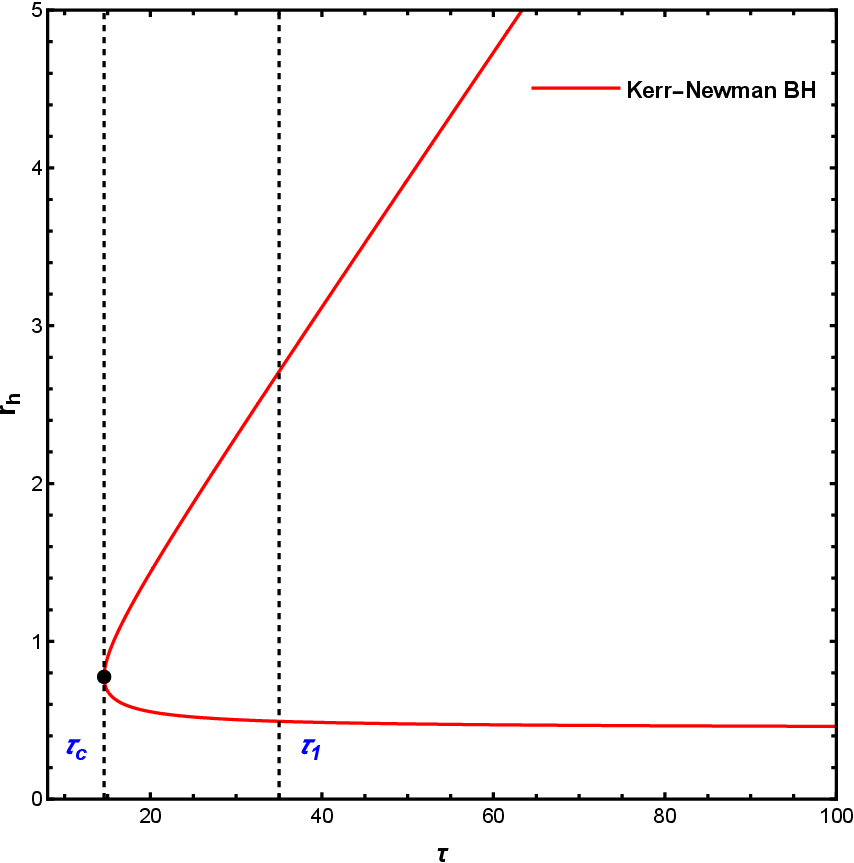}
}

\caption{
For a = 0.4 and Q = 0.2, the plot displays the behavior of the stable and unstable branches of the Kerr-Newman black hole phases as red curve.
}
\label{fig:OmegaVartheta}
\end{figure}

\begin{figure}[htbp]
\centering
\includegraphics[width=0.45\textwidth]{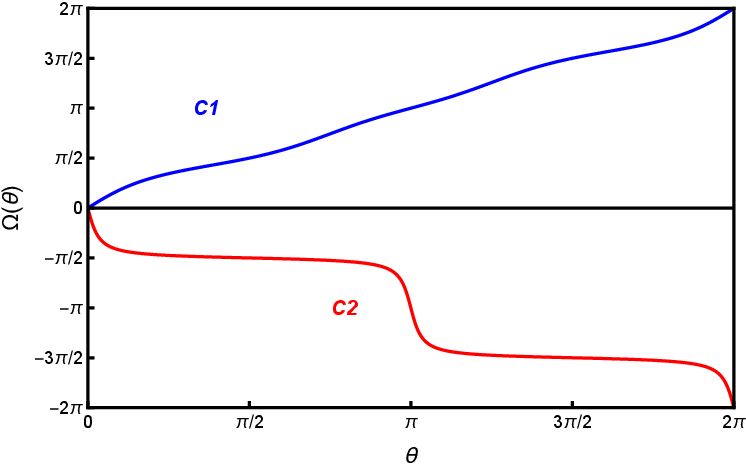}

\caption{The graph of $\Omega(\vartheta)$ against $\vartheta$ for the contours $C_2$ (Red solid curve) and $C_1$ (Blue solid curve) corresponding to the Kerr-Newman black hole with $\Lambda = 0$.}
\label{vector3}
\end{figure}
\begin{figure}[htbp]
\centering
\includegraphics[width=0.4\textwidth]{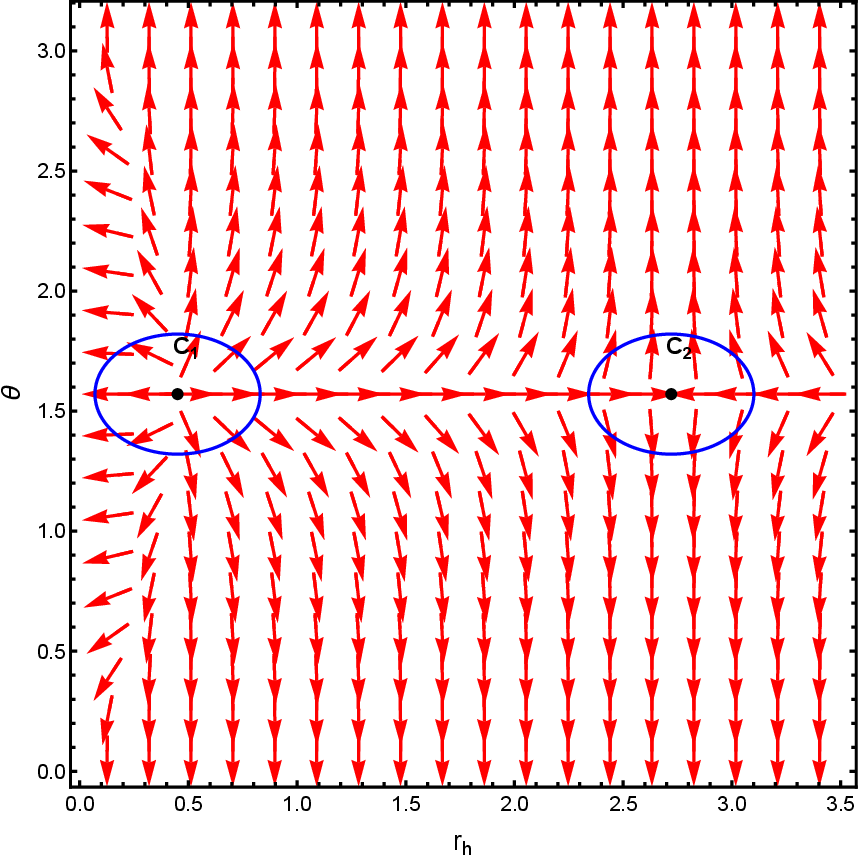}

\caption{A four-dimensional Kiselev black hole with rotation parameter a = 0.4, electric charge Q = 0.1, and $\tau=35$ has a unit vector field $n$ on the $r_h$--$\Theta$ plane. The zero points (ZPs), indicated by black dots, are situated at
$(r_h,\Theta)=(0.450319,\pi/2)$ and $(r_h,\Theta)=(2.72127,\pi/2)$.
The contours $C_1$ and $C_2$, enclosing the zero points, carry winding number
$w_1=+1$ and $w_2=-1$, respectively.
Their sum gives the total topological winding number $W=0$.}
\label{kiselev}
\end{figure}


 on contrast, we show the zero points of the component $\phi^{r_h}$ on the $(r_h-\tau)$ plane for the Kerr-Newman black hole with a set of rotation and electric charge parameter values of $a=0.4$ and $Q=0.2$ in Fig.\ref{fig:OmegaVartheta}. There are two Kerr-Newman black holes at $\tau=\tau_1$ and only one at $\tau=\tau_c$. The RN and Kerr black hole configurations, which describe the phase transition behavior of small to large black hole phases, are comparable to this behavior. The division of Kerr-Newman black holes into upper and lower branches, with corresponding winding numbers $w=-1$ and $w=+1$, is attributed to the creation point at $\tau=\tau_c$. \\
Plotting the  $\Omega(\vartheta)$  for the contours in Fig. \ref{vector3} illustrates how the topological number calculation works. Since the contours $C_1$ and $C_2$ do not overlap, the ZPs are topologically protected. Due to the opposing orientations of these contours, the winding numbers have the opposite signs. The positive topological charge is represented by the blue solid curve in Fig. \ref{vector3}, whereas the negative topological charge is represented by the red solid curve. Consequently, the total topological charge is zero.\\ 
We examine the limiting scenario where the spinning Kiselev geometry reduces to the Kerr–Newman black hole in order to verify the correctness of our thermodynamic topological approach. Under the identical parameter settings, we discover that the zero-point structure and accompanying winding numbers match those of the Kerr–Newman geometry. This correlation validates the Consistency of our findings and offers a foundation for differentiating the topological characteristics of the Kerr-Newman black hole from those resulting from the surrounding anisotropic matter.  Fig. \ref{kiselev} depicts the topological nature of the thermodynamic phase structure in the (\(r_h-\theta)\) plane of a four-dimensional Kiselev black hole with the ensemble temperature $\tau=35$ and $w=1/3$ in the off-shell configuration. The rotation and electric charge parameters are set at 0.4 and 0.1, respectively. The contours \(C_1\) and \(C_2\) are identified to be positioned at \((r_h, \Theta) = (0.450319, \pi/2)\) and \((2.72127, \pi/2)\), respectively. The orientations of the contours are such that \(C_1\) is oriented counterclockwise while \(C_2\) is oriented clockwise. The contours \(C_1\) and \(C_2\) are associated with topological charges of \(Q_{C_1} = +1\) and \(Q_{C_2} = -1\), respectively. This results in a total topological charge of \(Q_\text{total} = 0\). For your selected parameter values, the rotating Kiselev black hole displays the same topological charge as the Kerr-Newman black hole ($w=1/3$).

\section{Anisotropic Kerr-Newman-ADS topology employing complex residue method}
\label{III}
This part evaluates the topological number by the complex residue method \cite{fang2023revisiting}. This method requires us to determine the winding number ($w_i$) for each pole that comes from the residue method. The total topological charge is obtained by adding the winding numbers that correspond to the poles. Motivated by this, we first find the formula for $\tau$ as a function of the horizon radius by solving $\partial_{r_h}\mathcal{F}=0$.
\begin{equation}
\tau=\mathcal{G}(r_h) .
\end{equation} 
The horizon radius would therefore be substituted by the complex variable $z$, so that \cite{fang2023revisiting}$$\tau(r_h\to z)=\mathcal{G}(z)$$ 
When we replace the event horizon radius ($r_h$) with a complex variable $z$, it means that we have extended to the complex plane $\mathbb{C}$ with a complex number defined by $z=x+i y$, where $x, y\in \mathbb{R}$ are real variables. Our goal is to establish a two-dimensional parametric space where real contours can be specified. The equations of $\mathcal{G}(z)$ indicate the thermodynamic  defects that must be present in this contour.\\
The characterized complex function $\mathcal{R}(z)$ is then defined as follows
\begin{equation}\label{cf}
\mathcal{R}(z)\equiv\frac{1}{\tau-\mathcal{G}(z)} .
\end{equation}
The generalized free energy is associated with the  characterized  complex function $\mathcal{R}(z)$. With the exclusion of the singular point, the complex function $\mathcal{R}(z)$ is analytic on and inside the contour ${C_i}$. Using the residue theorem, the integral of the function $\mathcal{R}(z)$ along the contour ${C_i}$ surrounding $z_i$ can be found. According to the Cauchy-Goursat theorem, the integral of $\frac{\mathcal{R}(z)}{2\pi i}$ winding the singular points along the contour ${C}$ equals the integrals of $\mathcal{R}(z)$ along ${C_i}$ in the interior containing $z_i$.
\begin{figure}[htbp]
\centering
\includegraphics[width=0.5\textwidth]{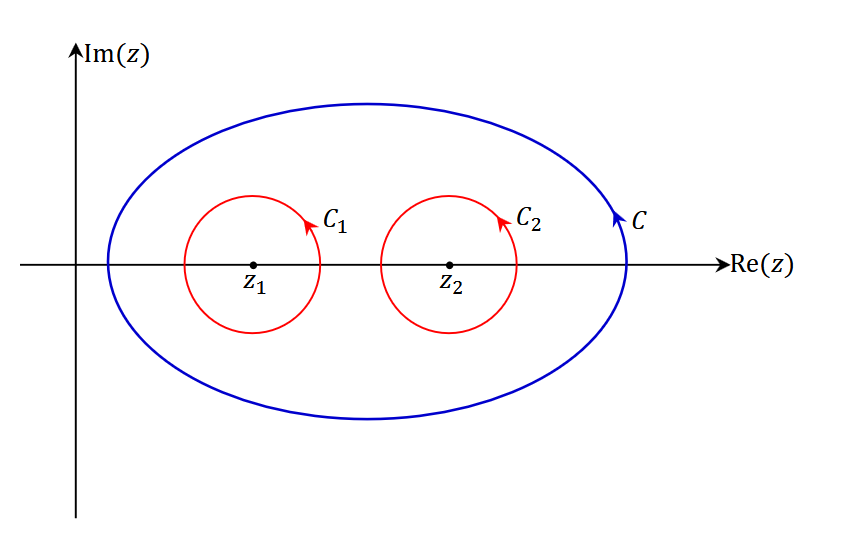}
\caption{Graphical illustration of the contour integral of a function along paths $C$ and $C_i$ that is analytic everyplace except at a finite isolated singular point $z_i$.}
\label{vector2}
\end{figure}

 According to complex residue theorem, the contour integral of the complex function \(R(z)\) is evaluated along the closed path \(C_k\), which surrounds the isolated singularity \(z_k\), and is expressed as

\begin{equation}
\oint_C \frac{\mathcal{R}(z)}{2\pi i} d z= \sum_{k=1}^m \operatorname{Res}\left[\mathcal{R}\left(z_k\right)\right] .
\end{equation}
As a result, the topological defects that were previously detected on the real plane by $\phi^{r_h}=0$ (as demonstrated in Eq. (\ref{phi}) now become isolated singular points of the described complex function $\mathcal{R}(z)$ on the $z$-plane.\\
An arbitrary closed contour can encircle each singular point, which represents a thermodynamic defect. With the exception of the isolated singularities \(z_1,z_2,\ldots,z_N\),the rational function \(\mathcal{R}(z)\) is analytic. The contour integral around a closed contour \(C\) that includes all singularities can be expressed using the residue theorem as the sum of the integrals around individual contours \(C_i\), each of which contains a singularity \(z_i\). 

This encourages us to reflect the winding number using the residue of the complex function at its pole point, which is also the topological defect corresponding to an on-shell black hole solution. We only need to normalize the residue to acquire the winding number in order to recover the topology information for the local defect spot. The winding number $w_i$ associated with a singular point $z=z_i$, is defined as \cite{fang2023revisiting}
\begin{equation}\label{wind}
w_i =\frac{\textrm{Res}\mathcal{R}(z_i)}{|\textrm{Res}\mathcal{R}(z_i)|}=\textrm{sgn}[\textrm{Res}\mathcal{R}(z_i)] .
\end{equation}
The expression $|\textrm{Res}\mathcal{R}(z_i)|$ refers to the absolute value of the complex function,i.e., $\textrm{Res}\mathcal{R}(z_i)$. The sign function, represented by the quantity $\textrm{sgn}(x)$, has the following functional characteristics
\[
\mathrm{sgn}(x)=
\begin{cases}
+1, & x>0 \\
0, & x=0 \\
-1, & x<0
\end{cases}
\]

Here, we should observe that we obtain $\textrm{Res}\mathcal{R}(z_i)\in\mathbb{R}$ since the singular points of concern are real. The black hole spacetime's global topological number $W$ can be computed as follows :
\begin{equation}\label{topw}
W=\sum_i w_i .
\end{equation}
This can be explained by the Cauchy-Goursat theorem, which enables us to compute an integral of the complex function around an exterior contour by integrals of simple interior closed contours including singular points. Thus, for the scenario depicted in Fig.12, we have
\begin{equation}
\label{caug}
\oint_C \mathcal{R}(z) d z= \sum_i \oint_{C_i} \mathcal{R}(z) d z .
\end{equation}

\begin{figure}[httb!]
\centering
\includegraphics[width=0.5\textwidth]{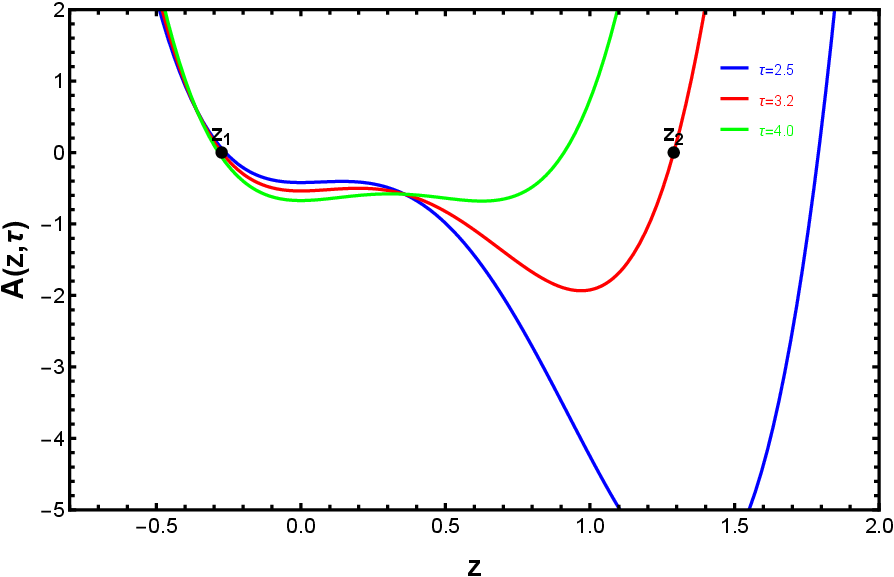}
\caption{roots of the polynomial equation $A(z)=0$ for various model parameter values. Variations in the parameter $\tau$, are represented by three curves. The evolution of the roots shows how the underlying physical characteristics affect the solution structure.}
\label{fig:roots_polynomial}
\end{figure}
We can deal with and identify the various types of isolated singular points by determining the thermodynamic defects winding number using residues.
In particular, this allows one to determine the number of windings in a pole of order greater than 1, this could be a creation point, annihilation point, or critical point. In the following, we shall calculate the thermodynamic topologies of typical black hole solutions of Einstein equations to demonstrate how these operate. In the present work, we explain how to compute the thermodynamic topologies of the Anisotropic Kerr-Newman-AdS black hole solutions.\\

The radial function for anisotropic Kerr-Newman AdS black hole is given by,
\begin{equation}
{
\Delta_r
=
(r^2+a^2)
\left(1+\frac{r^2}{L^2}\right)
+Q^2
-2Mr
-Kr^{2(1-\omega)}
} .
\label{eq:Delta_AdS}
\end{equation}
Equation (\ref{thermkerr}) yields the generalized off-shell free energy $\mathcal{F}$ of the anisotropic Kerr-Newman AdS black hole system as
\begin{equation}
\mathcal{F}(r_h,\tau)
=
\frac{
(r_h^2+a^2)
\left[
\tau\left(9+24\pi P r_h^2\right)
-6\pi r_h\left(3-8\pi P a^2\right)
\right]
+\tau Q^2\left(3-8\pi P a^2\right)^2
-9K\tau r_h^{2(1-\omega)}
}{
2r_h\tau\left(3-8\pi P a^2\right)^2
} .
\label{offFree_E1}
\end{equation}
To investigate the thermodynamic topology using the complex residue
approach, we introduce the characteristic rational function
\begin{equation}
R(z,\tau)
=
\frac{N(z)}{A(z,\tau)}
=
\frac{\Xi D(z)}
{\tau\Xi D(z)-4\pi z^3} .
\end{equation}

Here, \(D(z)\) is given by
\begin{equation}
D(z)=3z^4-0.16z^3+1.16z^2-0.20 .
\end{equation}

For the chosen parameter value \(\Xi=0.84\), the numerator becomes
\begin{align}
N(z)
&=\Xi D(z) \nonumber =2.52z^4-0.1344z^3+0.9744z^2-0.168 .
\end{align}

Accordingly, the denominator function can be written as
\begin{equation}
A(z,\tau)
=
2.52\tau z^4-(4\pi+0.1344\tau)z^3
+0.9744\tau z^2-0.168\tau .
\end{equation}

For \(\tau=3.2\), this reduces to
\begin{equation}
A(z,3.2)
=
8.064z^4-12.99645z^3
+3.11808z^2-0.5376 .
\end{equation}

The pole locations of the characteristic function are subsequently
determined from the roots of
\begin{equation}
A(z,3.2)=0 .
\end{equation}

This polynomial must be solved by setting it to zero for different parameter values. After finding z, we may use Eq.(\ref{wind}) to get the winding numbers. \\
We illustrate in Fig.~\ref{fig:roots_polynomial} the complex polynomial $\mathcal{A}(z)$ plot for three distinct temperatures. The complex residue analysis yields two real poles for the anisotropic Kerr--Newman-AdS black hole,
\[
z_1=-0.272236,
\qquad
z_2=1.35276 .
\]
However, since the event horizon radius must be real and positive
(\(r_h>0\)), the negative root \(z_1\) does not correspond to a
physical black hole horizon and is therefore excluded from the physical
thermodynamic analysis. Consequently, only the positive real pole
\(z_2=1.35276\) is considered as the physically relevant thermodynamic
defect.

Evaluating the residue of the characteristic function at \(z=z_2\)
yields a positive value. Consequently, the corresponding local winding
number is
\[
w_2
=
\operatorname{sgn}\!\left[\operatorname{Res}(R,z_2)\right]
=
+1 
\] 
Since only one physically relevant thermodynamic defect is present for
the chosen set of parameters, the total topological number is
\[
W=+1
\]
Therefore, the complex residue method characterizes the anisotropic
Kerr--Newman--AdS black hole by a positive thermodynamic winding number.
This result is consistent with the topological characterization obtained
from Duan's topological current theory.
\section{Conclusion}
\label{IV}

A powerful framework for examining the thermodynamic characteristics of black holes is topology. One can categorize various black hole configurations according to their topological characteristics by considering black hole phases in an appropriate thermodynamic parameter space as topological defects. This reveals universal characteristics that are independent of coordinate selection. Because it combines spin, electric charge, a negative cosmological constant, and an anisotropic matter field, the anisotropic Kerr–Newman–AdS black hole offers a very rich framework for investigating these concepts. The interaction of these components may result in unique topological characteristics and can drastically alter the thermodynamic phase structure. Inspired by this, we study the effects on the thermodynamic topology of the anisotropic Kerr–Newman–AdS black hole of rotation, electric charge, AdS pressure, and the anisotropic matter contribution. To investigate whether the topological properties of simpler black hole geometries are maintained or altered in this more general spacetime, we specifically analyze the associated topological defects, determine the global topological number and confirm their winding numbers.

In this paper, we study the thermodynamic topology of the anisotropic Kerr–Newman–AdS black hole using an independent complex-residue method and Duan's $\phi$-mapping theory. The topological defects and their winding numbers are found by constructing the relevant thermodynamic vector field in the $(r_h,\Theta)$ parameter space using the off-shell free energy within the $\phi$-mapping framework. First, we study the whole anisotropic Kerr–Newman–AdS configuration, including its topological structure and $r_h-\tau$ thermodynamic connection. Then, in order to separate the impacts of the AdS background and the anisotropic matter, we rigorously investigate two limiting instances. The Kerr–Newman–AdS black hole, for which the $r_h-\tau$ relation, vector field, and winding numbers are examined using the same framework, is obtained by setting $K=0$, which eliminates the anisotropic contribution. The Kerr–Newman black hole, whose thermodynamic topology is examined in a similar way, results from further setting the AdS contribution to zero along with $K=0$. In order to ascertain how the anisotropic matter and AdS background affect the topological categorization of the black holes, the winding numbers and global topological numbers obtained for these three configurations are then compared. Lastly, we independently compute the global topological number corresponding to the anisotropic Kerr-Newman-AdS scenario using the complex residue approach. Further evidence for the underlying topological structure of the anisotropic Kerr-Newman-AdS black hole is provided by the consistency between the two methods, which also establishes the robustness of the topological categorization.

A number of interesting aspects of the thermodynamic topology are revealed by examining the limiting cases. We identify a single thermodynamic defect with winding number $+1$ regarding the anisotropic Kerr--Newman--AdS black hole, leading to a  global topological number $W=+1$. The system reduces to the Kerr--Newman--AdS black hole when the anisotropic parameter is set to $K=0$. Interestingly, this configuration also possesses a single defect with winding number $+1$, which leads to $W=+1$. Consequently, the global topological number of the Kerr--Newman--AdS black hole remains unaffected by the addition of the anisotropic matter contribution. Additionally, we recover the Kerr--Newman black hole by setting $K=0$ and eliminating the AdS contribution. In this case, two distinct thermodynamic defects with winding numbers $+1$ and $-1$, respectively, emerge. Their contributions cancel each other, resulting in a total topological number $W=0$. These findings show that the global thermodynamic topology is significantly influenced by the AdS background, which causes the total topological number for the Kerr-Newman black hole to change from $W=0$ to $W=+1$ with respect to the Kerr-Newman-AdS configuration. On the other hand, the anisotropic matter contribution taken into consideration here maintains the Kerr-Newman-AdS black hole's topological class $W=+1$. This comparison demonstrates that the global topological features are sensitive to the AdS structure but resistant to the anisotropic matter field. Furthermore, reducing the rotating Kiselev geometry to the Kerr-Newman black hole reproduces the previously known thermodynamic topological results, so verifying our technique. This connection improves the consistency of our study and offers a foundation for discovering the topological effects caused by anisotropic matter in more general black hole configurations, increasing the relevance of our findings.
We also use the complex residue approach for the anisotropic Kerr–Newman–AdS black hole as an independent confirmation of these findings. We create the complex function \(R(z)\), whose unique points match the thermodynamic flaws, by analytically extending the pertinent thermodynamic function into the complex plane. The corresponding winding numbers are determined by the residues calculated at these single places through $w_i = \operatorname{sgn}\left[\operatorname{Res} R(z_i)\right]$. The residue analysis corresponding to anisotropic Kerr-Newman-AdS black hole produces the same total winding number $W=+1$ as Duan's $\phi$-mapping method. This agreement shows that the complex residue technique is consistent with the thermodynamic topology derived from the vector field analysis and offers an independent confirmation of topological classification.

Applying the complex residue approach, we further investigate thermodynamic topology of the anisotropic Kerr--Newman--AdS black hole. We create the equivalent complex polynomial $\mathcal{A}(z)$ and examine its roots for three distinct values of the temperature $\tau$ by extending the pertinent thermodynamic function to the complex plane. The distribution of the polynomial roots reflects the different thermodynamic regimes, which varies with temperature. Specifically, the relevant thermodynamic flaws and their winding numbers can be inferred from the number and location of the roots. The winding numbers derived from the complex residue approach are found to be consistent with those derived from Duan's $\phi$-mapping theory for the three representative temperatures taken into consideration. This agreement shows that the topological structure derived from the vector-field analysis is successfully reproduced by the complex residue approach. It also shows that the thermodynamic defects are directly represented mathematically by the roots of the complex polynomial, and the accompanying winding numbers are encoded by their associated residues. Thus, the complex residue technique can independently and complementarily characterize the thermodynamic topology of the anisotropic Kerr--Newman--AdS black hole.

This paper wraps up by illustrating that topological techniques offer a helpful framework for comprehending the anisotropic Kerr--Newman--AdS black hole's thermodynamic phase structure. Its topological faults and winding numbers are revealed by Duan's $\phi$-mapping theory, and the results are independently verified by the complex residue approach. The relevance of the AdS background and anisotropic matter in thermodynamic topology is further highlighted by the comparison with Kerr--Newman--AdS and Kerr--Newman black holes. These findings show the topological classification's resilience and could offer helpful information for upcoming research on quantum gravity and black hole thermodynamics.\\

\section*{Acknowledgement}
Author C.L. Ahmed Rizwan acknowledges the project grant received under the PM USHA Scheme G.O. number G.O(Rt)No.239/2025/HEDN dated 22.02.2025. Author M.S.A acknowledges the Inter-University-Centre for Astronomy and Astrophysics(IUCAA),Pune, India for providing research facilities and support under the Visiting Associateship Programme  

\bibliography{References.bib}
\bibliographystyle{unsrt}

\end{document}